\documentclass[12pt]{article} 
\usepackage[sectionbib]{natbib}
\usepackage{array,epsfig,fancyheadings,rotating}
\usepackage[]{hyperref}  
\usepackage{xcolor}
\usepackage{sectsty, secdot}
\sectionfont{\fontsize{12}{14pt plus.8pt minus .6pt}\selectfont}
\renewcommand{\theequation}{\thesection\arabic{equation}}
\subsectionfont{\fontsize{12}{14pt plus.8pt minus .6pt}\selectfont}

\usepackage{xr}
\usepackage{mathtools}
\usepackage{amsmath}
\usepackage{amssymb}
\usepackage{amsfonts}
\usepackage{multirow}
\usepackage{amsthm}
\usepackage{subfigure}
\usepackage{float}
\usepackage{alphalph}

\newtheorem{theorem}{Theorem}

\newtheorem{theorem3}[theorem]{Theorem}
\newtheorem{theorem4}[theorem]{Theorem}
\newtheorem{theorem5}[theorem]{Theorem}

\newtheorem{assumption1}{Assumption}
\newtheorem{assumption2}[assumption1]{Assumption}
\newtheorem{assumption3}[assumption1]{Assumption}
\newtheorem{assumption4}[assumption1]{Assumption}
\newtheorem{assumption5}[assumption1]{Assumption}

\theoremstyle{definition}

\newtheorem*{notation}{Notation}

\newtheorem{remark}{Remark}
\newtheorem{remark2}[remark]{Remark}
\makeatletter
\newcommand*{\addFileDependency}[1]{
  \typeout{(#1)}
  \@addtofilelist{#1}
  \IfFileExists{#1}{}{\typeout{No file #1.}}
}
\makeatother
\newcommand*{\myexternaldocument}[1]{%
    \externaldocument{#1}%
    \addFileDependency{#1.tex}%
    \addFileDependency{#1.aux}%
}
\myexternaldocument{supp-temp}

\begin{document}


\renewcommand{\baselinestretch}{2}

\markright{ \hbox{\footnotesize\rm Statistica Sinica
}\hfill\\[-13pt]
\hbox{\footnotesize\rm
}\hfill }

\markboth{\hfill{\footnotesize\rm FIRSTNAME1 LASTNAME1 AND FIRSTNAME2 LASTNAME2} \hfill}
{\hfill {\footnotesize\rm FILL IN A SHORT RUNNING TITLE} \hfill}

\renewcommand{\thefootnote}{}
$\ $\par


\fontsize{12}{14pt plus.8pt minus .6pt}\selectfont \vspace{0.8pc}
\centerline{\large\bf A Functional Coefficients Network Autoregressive Model}
\vspace{2pt} 
\centerline{\large\bf }
\vspace{.4cm} 
\centerline{Hang Yin, Abolfazl Safikhani and George Michailidis} 
\vspace{.4cm} 
\centerline{\it University of Florida, George Mason University and University of California, Los Angeles}
 \vspace{.55cm} \fontsize{9}{11.5pt plus.8pt minus.6pt}\selectfont


\begin{quotation}
\noindent {\it Abstract:}
The paper introduces a flexible model for the analysis of multivariate nonlinear time series data. The proposed Functional Coefficients Network Autoregressive (FCNAR) model considers the response of each node in the network to depend in a \textit{nonlinear} fashion to each own past values (autoregressive component), as well as past values of each neighbor (network component). Key issues of model stability/stationarity, together with model parameter identifiability, estimation and inference are addressed for error processes that can be heavier than Gaussian for both fixed and growing number of network nodes. The performance of the estimators for the FCNAR model is assessed on synthetic data and the applicability of the model is illustrated on two data sets; the first on multiple indicators of air pollution data and the second on Covid-19 cases in Florida counties.

\vspace{9pt}
\par
\end{quotation}\par

\def\thefigure{\arabic{figure}}
\def\thetable{\arabic{table}}

\renewcommand{\theequation}{\thesection.\arabic{equation}}

\fontsize{12}{14pt plus.8pt minus .6pt}\selectfont

\section{Introduction}\label{sec:intro}

Nonlinear time series models
gained prominence because of their ability to model, analyze and predict complex patterns in data in a wide range of fields, including economics and finance \cite{franses2000non}, climate \cite{donges2015unified}, cognitive science \cite{ward2002dynamical}, geosciences \cite{donner2008nonlinear} and engineering \cite{zou2019complex}. Departure from linearity opens different possibilities of developing nonlinear models.  However, a fully 
nonparametric model that does not impose any constraints on the autoregressive form becomes harder to estimate with limited data or in a multivariate setting \citep{fan2003nonlinear}. Hence, the literature focused on specific classes of \textit{parametric} nonlinear models for the \textit{conditional mean} for \textit{univariate} time series data, such as the exponential autoregressive (EXPAR) \citep{haggan1981modelling}, the threshold autoregressive  (TAR)  \citep{tong1990non} and the smooth transition autoregressive models \citep{dijk2002smooth} that added modeling flexibility.
The Functional Coeffecient Autoregressive Model (FAR) introduced by \citet{chen1993functional} encompassed these other classes of models and ergodicity, estimation and inference issues were addressed. Follow-up work by \citet{huang2004functional, chen2001functional} and \citet{cao2010penalized} proposed alternative estimation procedures for FAR models, as well as tests of whether the functional form of the FAR model is constant and hence it reduces to a linear one.

Note that a different line of work focused on nonlinear models for the \textit{conditional variance}, including nonlinear ARCH \citep{higgins1992class}, nonlinear thresholded ARCH \citep{gourieroux1992qualitative} and nonlinear GARCH \citep{hamadeh2011asymptotic} models.

In many applications, one is interested in modeling a large number of time series, whose interrelations are reflected through a \textit{network structure}. Motivated by social network related applications,
\citet{zhu2017network} developed a network vector autoregression (NAR) that 
captures the temporal dependence for the time series of each network node through its own history, as well as through that of neighboring nodes. A general form that encompasses various extensions of the basic NAR model for numerical variables was studied in \citet{yin2021general}, while \citet{armillotta2021poisson} considered a basic NAR model for Poisson count data.

Next, we present the general form of the NAR model presented in \cite{yin2021general}.
Consider a network comprising of $N$ nodes, for which measurements have been collected for $T$ time periods for a variable of interest $X$; i.e. $X_{it}, i=1,\cdots,N,\ t=1,\cdots,T$. 
Assuming lag-1 temporal dependence for ease of presentation, the model is given by
\begin{equation}\label{eq:NAR-defn}
X_{i,t}=a_i X_{i,(t-1)} + b_i w_i^T \mathbb{X}_{t-1}+\gamma_i^T Y_{i,(t-1)} +
\epsilon_{i,t},
\end{equation}
where $a_i, b_i \in\mathbb{R}, \  \gamma_i\in\mathbb{R}^p$ are regression coefficients, $Y_{i,(t-1)}$ a $p$-dimensional time-varying covariate vector, and $w_i$ a $N$-dimensional weight vector comprising of non-negative elements and summing to 1.
Hence, in the NAR model, measurements $X_{it}$ for node $i$ are influenced by their own past values, plus past values of combinations of ``related" nodes (network lags), plus \textit{time-varying} covariates. Constrained versions of the NAR model (e.g., assuming that all autoregressive/network effects are common across all nodes, i.e. $a_i=a$ and/or $b_i=b$ for all or selected groups of nodes $i=1,\cdots,N$ are special cases of the model in \eqref{eq:NAR-defn}; for details see \citet{yin2021general}.

\begin{figure} [H]
\centering
\subfigure[$a_i(\cdot)$]{
\includegraphics[scale=0.24]{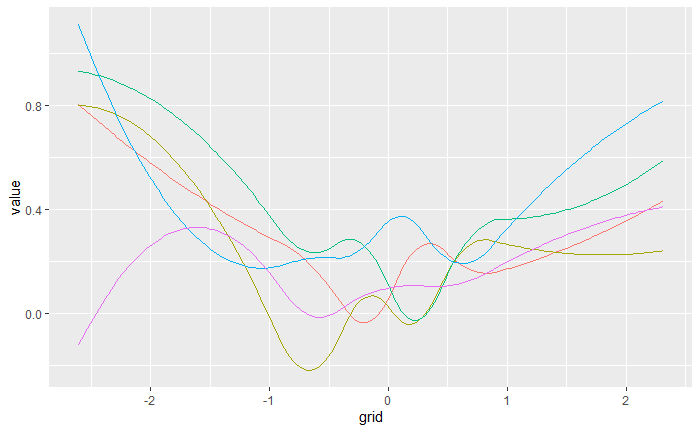}
}
\subfigure[$b_i(\cdot)$]{
\includegraphics[scale=0.24]{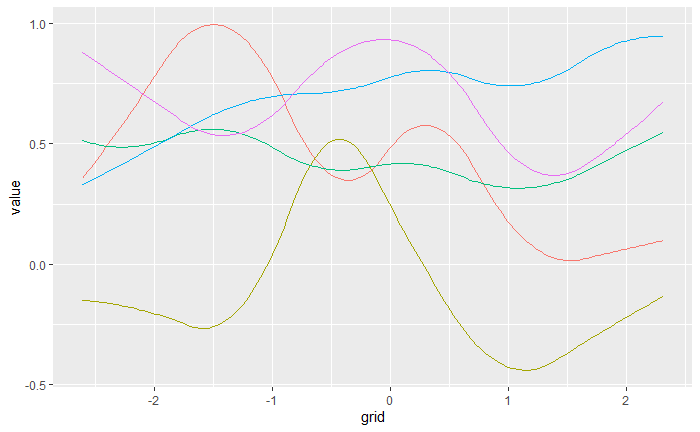}
} 
\caption{Plot of $a_i(\cdot)$ and $b_i(\cdot)$ with nonlinear patterns\label{fig:functions-of-pm10}}
\end{figure}

However, analogously to the univariate FAR model, both autoregressive and network effect coefficients $a_i/b_i$ can be \textit{time-varying}. Consider the daily PM10\footnote{Airborne particulate matter pollutants smaller in diameter than 10 microns} air quality indicator data, discussed in detail in Section \ref{sec:app}. As can be seen 
in Figure \ref{fig:functions-of-pm10}, the autoregressive and network coefficient functions for selected monitoring stations exhibit strong nonlinear patterns.

To accommodate such time-varying behavior, we introduce a Functional Coefficients Network Autoregressive (FCNAR) model described next. The FCNAR($q_1$,$q_2$) model with $q_1$ autoregressive lags and $q_2$ network lags takes the form 
\begin{equation}\label{e1}X_{it}=\sum\limits_{j=1}^{q_1}a_{i,j}(U_{it})X_{i,(t-j)}+\sum\limits_{j=1}^{q_2}b_{i,j}(U_{it})w_i^T\mathbb{X}_{t-j}+\epsilon_{it},
\end{equation}
where $U_{it}$ denotes a \textit{threshold} variable for node $i$, and can be either \textit{exogenously} determined or can correspond to lagged values of the outcome variable $X_{it}$. Hence, for node $i$ and lag $j$, $a_{i,j}(\cdot)$ and $b_{i,j}(\cdot)$ are unknown functions of the threshold variables $U_{it}$. 

%

The presence of the threshold variables $U_{it}$ introduces a number of technical challenges that relate to (i) the stability of the FCNAR process, (ii) the identifiability of the autoregressive and network effect parameters and (iii) the asymptotics for the $a_i(\cdot)/b_i(\cdot)$ functions and the FCNAR model parameters that require careful handling and resolved in the sequel.


The remainder of the paper is organized as follows. Section \ref{sec:stationary} discusses conditions to ensure the stability of the FCNAR process. Section \ref{sec:est} addresses model estimation, inference and testing issues, while Section \ref{sec:pe} presents results from synthetic data showcasing the performance of the developed estimators in different simulation settings. Section \ref{sec:app} uses the FCNAR model to analyze the followoing two data sets: (i) air pollution indicators and (ii) new Covid-19 cases at the county level in the state of Florida. Finally, some concluding remarks are summarized in Section \ref{sec:con}.

\begin{notation}
 Throughout the paper, $\mathbb{R}$ denotes the set of real numbers, respectively. We use $||A||$ and $||A||_F$ to denote the operator norm and Frobenius norm of a matrix $A$, $\oplus$ the direct sum of two matrices, and $\otimes$ the Kronecker product of two matrices of appropriate dimensions, $\text{diag}(A)$ a diagonal matrix $A$ and $a^T/A^T$ the transpose of a vector/matrix, respectively. For matrices, we use $\rightarrow_p$ to denote element-wise convergence in probability, and $\rightarrow_d$ to denote convergence in distribution. For a symmetric or Hermitian matrix $A$, we denote its spectral radius by $\rho(A)$. 
 \end{notation}

\section{Stability of the FCNAR process }\label{sec:stationary}

The first issue addressed is to derive the stability condition of the $\text{FCNAR}(q_1,q_2)$ process for the model defined in \eqref{e1}. To proceed, we rewrite equation \eqref{e1} in matrix form: 
\begin{equation}\label{e2}
\mathbb{X}_{t}=\sum\limits_{j=1}^{q_1}A_j(\mathbb{U}_{t})\mathbb{X}_{t-j}+\sum\limits_{j=1}^{q_2}B_j(\mathbb{U}_{t})W\mathbb{X}_{t-j}+\epsilon_t,
\end{equation}
where $\mathbb{U}_{t}$ is the threshold variable, $A_j(\mathbb{U}_{t}):=\text{diag}\{a_{1,j}(U_{1,(t-1)})\cdots a_{N,j}(U_{N,(t-1)})\}$, $B_j(\mathbb{U}_{t}):=\text{diag}\{b_{1,j}(U_{1,(t-1)})\cdots b_{N,j}(U_{N,(t-1)})\}$ and $W$ denotes the network weight matrix.

Define $G_j(\mathbb{U}_{t}):=A_j(\mathbb{U}_{t})+B_j(\mathbb{U}_{t})W.$ We can rewrite \eqref{e2} as $
\mathbb{X}_t=\sum\limits_{j=1}^{q}G_j(\mathbb{U}_{t})\mathbb{X}_{t-j}+\epsilon_{t}
$, wherein $j=1,2,\cdots,\max\{q_1,q_2\}$, with the convention that zero matrices are included/padded for the relationship to hold; namely, if $q_1>q_2$, $B_j(\mathbb{U}_{t})=0$ for $j>q_2$, whereas if $q_1<q_2$, $A_j(\mathbb{U}_{t})=0$ for $j>q_1$. Let $q=\max\{q_1,q_2\}$. Note that the model \eqref{e2} can also be expressed as a Vector Autoregressive model with a single lag (VAR(1)) \citep{lutkepohl2005new} as:
\begin{equation}
\label{var1}
\boldsymbol{X}_t=\boldsymbol{G}(\mathbb{U}_{t})\boldsymbol{X}_{t-1}+\mathcal{E}_t, \\ \ \ \text{with}
\end{equation}
\begin{equation}
\label{XEG}
\begin{split}
&\boldsymbol{X}_t:=\begin{bsmallmatrix}\mathbb{X}_t^T\\\mathbb{X}_{t-1}^T\\\vdots\\\mathbb{X}_{t-q+1}^T\end{bsmallmatrix},\ \mathcal{E}_t:=\begin{bsmallmatrix}\epsilon_t^T\\0^T\\\vdots\\0^T \end{bsmallmatrix}, \boldsymbol{G}(\mathbb{U}_{t}):=\begin{bsmallmatrix}G_1(\mathbb{U}_{t})&\cdots&G_{q-1}(\mathbb{U}_{t})&G_q(\mathbb{U}_{t})\\I_N&\cdots&0&0\\\vdots&\ddots&\vdots&\vdots\\0&\cdots&I_N&0 \end{bsmallmatrix}.
\end{split}
\end{equation}

Next, we introduce the needed assumptions to establish the stability result.
\begin{assumption1}\label{a1} The autorgressive
$a_{i,j}(\cdot)$ and network effect $b_{i,j}(\cdot)$ functions are bounded by universal constants $\tilde{a}_{i,j}$ and $\tilde{b}_{i,j}$ for all $i=1,2,\cdots,N$ and $j=1,2,\cdots,q$.
\end{assumption1}
\begin{assumption2}\label{a2}
The error process
       $\{\epsilon_t,t\in \mathbb{N}\}$ is a sequence of independent and identically distributed (iid) random vectors with $\mathbb{E}(\epsilon_t)=0$. Further, the marginal distribution of $\epsilon_{t}$ is absolutely continuous with respect to the Lebesgue measure and its density function is positive everywhere on $\mathbb{R}^N$.
    \end{assumption2}

Next, define $\tilde{A}_{j}:=\text{diag}\{\tilde{a}_{1,j},\tilde{a}_{2,j},\cdots,\tilde{a}_{N,j}\}$, $\tilde{B}_{j}:=\text{diag}\{\tilde{b}_{1,j},\tilde{b}_{2,j},\cdots,\tilde{b}_{N,j}\}$, and hence $\tilde{G}_j:=\tilde{A}_j+\tilde{B}_jW$, Then, due to Assumption 1, $\boldsymbol{G}(\cdot)$ is elementwise bounded by $\boldsymbol{\tilde{G}}$ where \[\boldsymbol{\tilde{G}}:=\begin{bsmallmatrix}\tilde{G}_1&\cdots&\tilde{G}_{q-1}&\tilde{G}_q\\I_N&\cdots&0&0\\\vdots&\ddots&\vdots&\vdots\\0&\cdots&I_N&0 \end{bsmallmatrix}.\]

We can then establish the following result.
\begin{theorem}
\label{t:ergodic}
Suppose Assumptions~\ref{a1}-\ref{a2} hold. Then, the FCNAR$(q_1,q_2)$ process defined in \eqref{e1} is geometrically ergodic (and hence stable), if $\rho(\boldsymbol{\tilde{G}})<1$.
\end{theorem}

\begin{remark}\label{impact-W-spectral-radius}
The density (number of neighbors) of the weight matrix $W$ impacts the structure of $\tilde{G}$ and thus of the spectral radius $\rho(\tilde{G})$. Numerical evidence (presented in the Supplement, Section \ref{impact-W}) indicates that the value of the spectral radius gradually decreases, once the bandwidth (number of neighbors) of the weight matrix $W$ exceeds a certain threshold. 
\end{remark}

\section{Estimation of the FCNAR model parameters and their Asymptotic Properties}
\label{sec:est}

\subsection{Estimation Procedure}

As mentioned in the introductory section, the functions of univariate FAR models can be estimated by various approaches, including using local polynomial and locally constant methods, local linear regression techniques, as well as polynomial and penalized splines methods.

In this work, for estimating the unknown functions $a_{i,j}(\cdot)$ and $b_{i,j}(\cdot)$ of the FCNAR model, we use splines. Further, for ease of presentation, we assume that $q_1=q_2=1$. The extension to general $q_1, q_2$ is straightforward.

For the order-M spline constructed by the truncated-power basis set, define the spline basis for $a_{i,j}(\cdot)$ and $b_{i,j}(\cdot)$ to be $\Phi_{i}(u)$, where:
\begin{equation}\label{bi1}
\Phi_{i}(u)=\begin{bsmallmatrix}\phi_{i1}(u)&\phi_{i12}(u)&\cdots&\phi_{i(M+K)}(u) \end{bsmallmatrix}
\end{equation}
where $ \phi_{i1}(u):=1,\ \phi_{i2}(u):=u,\ \phi_{i3}(u):=u^2,\cdots,\ \phi_{iM}(u):=u^{M-1},\\ \phi_{i(M+1)}(u):=(u-k_{1}^{(i)})_{+}^{M-1},\ \cdots,\ \phi_{i(M+K)}(u):=(u-k_K^{(i)})_{+}^{M-1}$.

We can then express the autoregressive and network effect coefficient functions as follows:
\begin{equation*}
\begin{split}
    a_{i,j}(u) =& a_{ij1}+\cdots a_{ijM} u^{M-1}+a_{ij(M+1)} (u-k_1^{(i)})_{+}^{M-1}+\cdots a_{ij(M+K)} (u-k_{K}^{(i)})_{+}^{M-1},
    \end{split}
\end{equation*}
\begin{equation*}
\begin{split}
    b_{i,j}(u) =& b_{ij1}+\cdots b_{ijM} u^{M-1}+b_{ij(M+1)} (u-k_1^{(i)})_{+}^{M-1}+\cdots b_{ij(M+K)} (u-k_{K}^{(i)})_{+}^{M-1},
    \end{split}
\end{equation*}
where $M$ denotes the order of the spline and $\{k_j^{(i)}\}_{j=1}^K$ the sequence of spline knots for node $i$. Hence, estimating the autoregressive $a_{i,j}(\cdot)$ and networ effect $b_{i,j}(\cdot)$ coefficient functions is equivalent to estimating the parameters $a_{ij1}, a_{ij2}\cdots, a_{ij(M+K)}$ and $b_{ij1}, b_{ij2}\cdots, b_{ij(M+K)}$ for the corresponding spline bases. 

In the sequel, we consider both ordinary least squares and ridge estimators for the spline base parameters. 
Defining \[Z_{i1}(X_{i(t-j)}, U_{it}):=\begin{bsmallmatrix}
    \phi_{i1}(U_{it})X_{i(t-j)}&\phi_{i2}(U_{it})X_{i(t-j)}&\cdots&\phi_{i(M+K)}(U_{it})X_{i(t-j)}
    \end{bsmallmatrix},\]
    \[Z_{i2}(w_i^T\mathbb{X}_{t-j},U_{it}):=\begin{bsmallmatrix}
    \phi_{i1}(U_{it})w_i^T\mathbb{X}_{t-j}&\phi_{i2}(U_{it})w_i^T\mathbb{X}_{t-j}&\cdots&\phi_{i(M+K)}(U_{it})w_i^T\mathbb{X}_{t-j}
    \end{bsmallmatrix},\] \[Z_{i(t-j)}=\begin{bmatrix}Z_{i1}(X_{i(t-j)}, U_{it})&Z_{i2}(w_i^T\mathbb{X}_{t-j},U_{it})\end{bmatrix},\]  \[\mathbb{Z}_{i(t-1)}:=\begin{bsmallmatrix}
    Z_{i(t-1)}&Z_{i(t-2)}&\cdots&Z_{i(t-q)}\\
\end{bsmallmatrix},\]

the model can then be written in compact form as:
\begin{equation}\label{compact}
    \mathbb{X}_{t}=\mathbb{Z}_{t-1}\beta+\epsilon_t,
\end{equation}
where 
\[\mathbb{Z}_{t-1}:=\begin{bsmallmatrix}
    \mathbb{Z}_{1(t-1)}&0&\cdots &0\\
    0&\mathbb{Z}_{2(t-1)}&\cdots&0\\
    \vdots&\vdots&\ddots&\vdots\\
    0&0&\cdots&\mathbb{Z}_{N(t-1)}\\
\end{bsmallmatrix},\]
\[\beta:=\begin{bsmallmatrix}\beta_{1}^T&\beta_2^T&\cdots&\beta_N^T\end{bsmallmatrix}^T,\ \beta_i:=\begin{bsmallmatrix}\beta_{i1}^T&\beta_{i2}^T&\cdots&\beta_{iq}^T\end{bsmallmatrix}^T, \textit{ and }\] \[\beta_{ij}:=\begin{bsmallmatrix}a_{ij1}&a_{ij2}&\cdots&a_{ij(M+K)}&b_{ij1}&b_{ij2}&\cdots&b_{ij(M+K)}\end{bsmallmatrix}^T.\]

Then, the least squares (LS) estimator is given by:
\[\hat{\beta}=(\sum\limits_{t=1}^T\mathbb{Z}_{t-1}^T\mathbb{Z}_{t-1})^{-1}\sum\limits_{t=1}^T\mathbb{Z}_{t-1}^T\mathbb{X}_{t}.\]

Note that the number of knots $K$, the placement of knots and the spline order $M$ are critical tuning parameters that impact the quality of the LS estimator in empirical work and are investigated in detail in Section \ref{sec:pe}.


In the presence of many lags and a large network size, the columns of the design matrix $Z$ may be fairly strongly correlated. Hence, a ridge estimator may be preferable in empirical work. Further, \cite{cao2010penalized} mention that the number and the location of knots are no longer as critical to be carefully selected (unlike the LS estimator) and can be controlled by the tuning parameter $\lambda$ of the ridge estimator, the latter defined as:
\begin{equation}
 \label{beta-ols-r}   
 \hat{\beta}^{ridge}=(\sum\limits_{t=1}^T\mathbb{Z}_{t-1}^T\mathbb{Z}_{t-1}+\lambda T \Psi)^{-1}\sum\limits_{t=1}^T\mathbb{Z}_{t-1}^T\mathbb{X}_{t},
\end{equation}
where $\Psi$ is a diagonal matrix with its (1 + iM+iK)-th diagonal elements equal to 0 and the rest equal to 1 for $i=0,1,\cdots,2Nq-1$, which penalizes the higher order coefficients.

\subsection{Asymptotic Properties of the Estimators}

Next, we introduce the needed assumptions for the main results. 
  \begin{assumption3}\label{a3}  Assumptions on the network matrix $W$:
  \begin{enumerate}
    \item[(a)] The eigenvalues of $E(Y_{it}Y_{it}^T|U_{it}=u)$ are uniformly bounded away from both 0 and infinity for all $u\in\mathbb{R}$ and $i=1,2,\cdots,N$, where $Y_{it}:=\begin{bmatrix}X_{it}&w_i^T\mathbb{X}_t\end{bmatrix}^T$.
    \item[(b)] $W\in \mathbb{R}^{N\times N}$ is a row-normalized matrix; i.e., $\sum_{j=1}^{N} w_{ij}=1$ with $w_{ij}\geq0$.
  \end{enumerate}
\end{assumption3}
\begin{assumption4}\label{a4} Assumptions on $\mathbb{U}_t$:
\begin{enumerate}
    \item[(a)] The marginal density of $U_{it}$ is bounded away from both zero and infinity uniformly on $\mathbb{R}$. 
 \item[(b)] One of the following cases needs to hold:
    \begin{itemize}
        \item The threshold process $\mathbb{U}_{t}$ corresponds the lagged values of the outcome process $\mathbb{X}_{t}$, or
        \item $\mathbb{U}_{t}$ is an exogenous process (independent of $\mathbb{\epsilon}_{t})$ and strictly stationary. Further, for some sufficiently large $m>0$, $\mathbb{E}(|U_{it}|^m)<\infty$.
    \end{itemize}
    \end{enumerate}
\end{assumption4}
\begin{assumption5}\label{a5} The error process $\{\epsilon_t,t\in \mathbb{N}\}$ satisfies $\mathbb{E}(\epsilon_t)=0$, $\Sigma_\epsilon=\sigma^2I$, and
$\mathbb{E}(|\epsilon_{it}|^4)<\infty$.

\end{assumption5}

Assumptions \ref{a3}(a) and \ref{a4}(a) are needed to ensure the \textit{identifiability} of the autoregressive and network effect coefficient functions (see discussion in \cite{huang2004functional}). 
Further, assumptions \ref{a3}(b), \ref{a4}(b) and \ref{a5} are required for establishing the asymptotic properties of the LS and ridge estimators of the model parameters. Specifically, the moment conditions on $X_{it}$ and $U_{it}$ ensure the martingale central limit theorem. Note that \ref{a4}(b) is also used in univariate FAR models. Assumptions~\ref{a1}, \ref{a2} together with \ref{a4}(b) imply that 
$\{\mathbb{U}_{t},\mathbb{X}_{t}\}$ is jointly strictly stationary. Further, note that Assumption \ref{a2} ensures identifiability of the autoregressive ($q_1$) and network effect ($q_2$) lag-orders. It is also worth noting that this is a standard assumption used also in univariate autoregressive time series models \citep{hamilton2020time} to ensure identifiability of the lag-order.
Finally, note that since we assume that $\Sigma_\epsilon=\sigma^2I$, the FCNAR parameter $\beta$ can be estimated element-wise.

\begin{remark}\label{impact-W-assumption-3a}
Note that the density (number of neighbors) of the network matrix $W$ impact the maximum and minimum eigenvalue of $E(Y_{it}Y_{it}^T|U_{it}=u)$. Numerical evidence (presented in the Supplement, Section \ref{impact-W}) indicates that the both the maximum and minimum eigenvalues do not change, once the bandwidth (number of neighbors) of the network matrix $W$ is above a certain number. Further, the minimum eigenvalue clearly remains bounded away from zero.
\end{remark}

Hence, for both fixed and diverging (with $T$) network size $N$, we establish the following three results:

\begin{theorem}[Asymptotic Properties of the LS Estimator]\label{t:ols:s} 
Suppose Assumptions 1-5 hold. Then, the node parameters of the FCNAR model \eqref{compact}, i.e., $\mathbb{X}_{t}=\mathbb{Z}_{t-1}\beta+\epsilon_t$, satisfy as $T\rightarrow \infty$
    \[\sqrt{T}(\hat{\beta}_i-\beta_i)\rightarrow_d N(0, \sigma^2P_i^{-1}), \ \ i=1,\cdots,N,\]
where $P_i:=\mathbb{E}(\mathbb{Z}_{i(t-1)}^T\mathbb{Z}_{i(t-1)}).$
\end{theorem}

\begin{remark}
In practice, the quantity $\mathbb{E}(\mathbb{Z}_{i(t-1)}^T\mathbb{Z}_{i(t-1)})$ can be estimated by $\frac{1}{T}\sum\limits_{t=1}^T\mathbb{Z}_{i(t-1)}^T\mathbb{Z}_{i(t-1)}$. 
\end{remark}

\begin{remark}
Note that a \textit{uniform convergence result} over the domain of the autoregressive $a_i(\cdot)$ and network effect  $b_i(\cdot)$ functions is given in Section 0.2.1 of the Supplement. 
\end{remark}



\begin{theorem3}[Asymptotic Properties of the Ridge Estimator]\label{T:ridge}
    Suppose Assumptions 1-5 hold and further assume that $\lambda = o(\frac{1}{\sqrt{T}})$. Then, the regression coefficient of the FCNAR model satisfies, as $T\rightarrow \infty$
    \[\sqrt{T}(\hat{\beta}_i^{ridge}-\beta_i)\rightarrow_d N(0, \sigma^2P_i^{-1}),\]
where $P_i:=\mathbb{E}(\mathbb{Z}_{i(t-1)}^T\mathbb{Z}_{i(t-1)}).$
    \end{theorem3}




The next result provides the asymptotic distribution of any point in the domain of the autoregressive and network effect coefficient functions. Define \[\beta_i(u):= \begin{bsmallmatrix}a_{i,1}(u) & b_{i,1}(u) & \cdots&a_{i,q}(u)&b_{i,q}(u)\end{bsmallmatrix}^T\in\mathbb{R}^{2q}\] to be a vector of the unknown autoregressive and network effect coefficient functions, we then have that $\beta_i(u)=(I_{2q} \otimes\Phi_i(u))\beta_i$.
%
    
\begin{theorem4}\label{P:OLS}
    Suppose Assumptions 1-5 hold. 
     Let $\hat{\beta}_i(u)$ be vectors of either the LS or the ridge estimator. Further, for the ridge estimator, assume that $\lambda = o(\frac{1}{\sqrt{T}})$. Then, for $U_{it}=u$,
    \[\sqrt{T}(\hat{\beta}_i(u)-\beta_i(u))\rightarrow_d N(0, \sigma^2(I_{2q} \otimes\Phi_i(u))P_i^{-1}(I_{2q} \otimes\Phi_i(u))^T).\]

\end{theorem4}

\begin{remark2}
In practice, we use the AIC criterion to select the lags of $U_{it}$, the number of knots $K$ and the spline order $M$ (for a discussion, see \cite{huang2004functional}). 
\end{remark2}

Also, note that since we assume $\Sigma_\epsilon=\sigma^2I$, for diverging $N$ and $T$, we can get the joint asymptotic distribution of any sub-vector of the FCNAR parameter $\beta$ of fixed dimension. Suppose we are interested in a subset of nodes in $\mathbb{S}:=\{N_1,N_2,\cdots,N_{S}\}\subset \{1,2,\cdots,N\}$, which is of fixed cardinality $S$. Denote $\beta_{sv}:=\begin{bmatrix}\beta_{N_1}^T&\beta_{N_2}^T&\cdots&\beta_{N_S}^T\end{bmatrix}^T\in \mathbb{R}^{2S(M+K)q}$ to be the sub-vector of $\beta$. For example, if we are interested in the joint asymptotic distribution of $\beta_1$ and $\beta_2$, in this case, $\beta_{sv}=\begin{bmatrix}\beta_1^T&\beta_2^T\end{bmatrix}^T$. Similarly, define \[\beta_{sv}(u):= \begin{bsmallmatrix}\beta_{N_1}(u)^T&\beta_{N_2}(u)^T&\cdots&\beta_{N_S}(u)^T&\end{bsmallmatrix}^T\in\mathbb{R}^{2Sq},\] and $\Phi_{sv}(u):=\oplus_{i\in \mathbb{S}}I_{2q} \otimes\Phi_i(u)$. We then have that $\beta_{sv}(u)=\Phi_{sv}(u)\beta_{sv}$.

\begin{theorem5}\label{t:joint}
Suppose assumptions 1-5 hold. Then, for any sub-vector $\beta_{sv}\in \mathbb{R}^{2S(M+K)q}$ of $\beta$, the regression coefficient of the FCNAR model satisfies, as $T\rightarrow \infty$
    \[\sqrt{T}(\hat{\beta}_{sv}-\beta_{sv})\rightarrow_d N(0, \sigma^2P_{sv}^{-1}),\]
where $P_{sv}:=\oplus_{i\in \mathbb{S}}P_i.$

Further, assume that $\lambda = o(\frac{1}{\sqrt{T}})$. Then, the regression coefficient of the ridge satisfies, as $T\rightarrow \infty$
\[\sqrt{T}(\hat{\beta}_{sv}^{ridge}-\beta_{sv})\rightarrow_d N(0, \sigma^2P_{sv}^{-1}).\]

And for $U_{it}=u$,
    \[\sqrt{T}(\hat{\beta}_{sv}(u)-\beta_{sv}(u))\rightarrow_d N(0, \sigma^2\Phi_{sv}(u)P_{sv}^{-1}\Phi_{sv}(u)^T).\]
\end{theorem5}

\subsection{Hypothesis Testing in FCNAR Models}
\label{sec:testing}

The technical results established in Theorem \ref{t:ols:s} enable testing hypotheses of interest on a \textit{fixed} subset of network nodes, as discussed next. Recall that a hypothesis involving  $p$ linear constraints on $\beta$ can be written as:
\[H_0: \mathbb{D}\beta = r \ \ \text{vs} \ \
H_a:\mathbb{D}\beta \not= r,\]
where $\mathbb{D}$ for the FCNAR model is a $p\times (2NMq+2NKq)$ matrix and $r$ is a $p\times 1$ column vector. Then, the test statistic takes the form
$$F=((\mathbb{D}\hat{\beta})'(\hat{\sigma}^2\mathbb{D}(\sum_{t=1}^{T}\mathbb{Z}_{t-1}^T\mathbb{Z}_{t-1})^{-1}\mathbb{D}')^{-1}(\mathbb{D}\hat{\beta}))/p \sim F(p, N(T-2Mq-2Kq))$$
and follows an $F$-distribution with the provided degrees of freedom, where $\hat{\sigma}^2=\frac{1}{NT}\sum\limits_{i=1}^N\sum\limits_{t=1}^T(X_{it}-\mathbb{Z}_{i(t-1)}^T\hat{\beta}_i)^2$.

For the FCNAR model,  hypotheses of particular interest are whether the autoregressive effects $a_{ij}(u)$ or the network effects $b_{ij}(u)$ are the same among nodes, and take the following form:

\noindent(A) \textit{Homogeneity of the autoregressive effects amongst nodes}: \[ H_0:a_{ijk}=a_{i'jk},\ \ \ \ \forall \ k=1,2,\cdots,(M+K),\ \forall \ i,i'=1,2,\cdots,N\text{ and }i\not=i'.\]
(B) 
\textit{Homogeneity of network effects amongst the nodes}: \[ H_0:b_{ijk}=b_{i'jk},\ \ \ \ \forall \ k=1,2,\cdots,(M+K),\ \forall \ i,i'=1,2,\cdots,N\text{ and }i\not=i'.\]

Let $\boldsymbol{\tilde{0}}$ denote the zero matrix of dimension $(M+K)\times (2q-1)(M+K)$ and $\boldsymbol{\Bar{0}}$ the zero matrix of dimension $(M+K)\times (M+K)$. For the homogeneity test of the lag-1 autoregressive effect, the constraint matrix $\mathbb{D}$ takes the form
\[\mathbb{D}:=\begin{bsmallmatrix}
\begin{array}{cc|cc|cc|c|cc}
I_{M+K}&\boldsymbol{\tilde{0}}&-I_{M+K}&\boldsymbol{\tilde{0}} &\boldsymbol{\Bar{0}}&\boldsymbol{\tilde{0}}&\cdots&\boldsymbol{\Bar{0}}&\boldsymbol{\tilde{0}}\\
\boldsymbol{\Bar{0}}&\boldsymbol{\tilde{0}}&I_{M+K}&\boldsymbol{\tilde{0}}&-I_{M+K}&\boldsymbol{\tilde{0}}& \cdots&\boldsymbol{\Bar{0}}&\boldsymbol{\tilde{0}}\\
\vdots&\vdots&\vdots&\vdots&\vdots&\vdots&\cdots&\vdots&\vdots\\
\boldsymbol{\Bar{0}}&\boldsymbol{\tilde{0}}&\boldsymbol{\Bar{0}}&\boldsymbol{\tilde{0}}&\boldsymbol{\Bar{0}}&\boldsymbol{\tilde{0}}&\cdots&-I_{M+K}&\boldsymbol{\tilde{0}}\\
\end{array}
\end{bsmallmatrix}.\]

Another set of hypotheses of particular interest are those pertaining to the functions of the autoregressive effects and the network effects being constant, which translates that the corresponding relationship to the outcome variable is \textit{linear}. These hypotheses take the form:

\noindent(C1) \textit{Linearity of the autoregressive effects}: \[ H_0:a_{ij2}=a_{ij3}=\cdots=a_{ij(M+K)}=0.\]
(C2) \textit{Linearity of the network effects}: \[ H_0:b_{ij2}=b_{ij3}=\cdots=b_{ij(M+K)}=0.\]

If the null hypothesis in any of these two cases holds, it leads to a simplification in the FCNAR model specification and a reduction in the number of model parameters to be estimated. If both hypotheses (A) and (B) hold, the FCNAR model can be reduced to the NAR model.

For hypothesis (A) the constraint matrix
$\mathbb{D}$ takes the form \[\mathbb{D}:=\begin{bmatrix}
\begin{array}{ccccc|c}
0&1&0&\cdots&0&\boldsymbol{0}_{M+K}^T\\
0&0&1&\cdots&0&\boldsymbol{0}_{M+K}^T\\
\vdots&\vdots&\vdots&\vdots&\vdots&\vdots\\
0&0&0&\cdots&1&\boldsymbol{0}_{M+K}^T
\end{array}
\end{bmatrix}.\]
%
Then, the F-test statistic is given by:
\[F=((\mathbb{D}\hat{\beta}_{ij})'(\hat{\sigma}^2\mathbb{D}(\sum_{t=1}^{T}Z_{i(t-j)}^TZ_{i(t-j)})^{-1}\mathbb{D}')^{-1}(\mathbb{D}\hat{\beta}_{ij}))/(M+K-1)\]
has a F-distribution with $(M+K-1, T-2Mq-2Kq)$ degrees of freedom, and $\hat{\sigma}^2=\frac{1}{T}\sum\limits_{t=1}^T(X_{it}-\mathbb{Z}_{i(t-1)}^T\hat{\beta}_i)^2$.

Under the alternative hypothesis, the non-centrality parameter is given by $\delta=(\mathbb{D}\beta_{ij})'(\sigma^2\mathbb{D}(\sum_{t=1}^{T}Z_{i(t-j)}^TZ_{i(t-j)})^{-1})\mathbb{D}')^{-1}(\mathbb{D}\beta_{ij})$. Analogous expressions can be derived for testing hypothesis (B).

\section{Performance Evaluation of the LS and Ridge Estimators}
\label{sec:pe}

We examine the performance of the LS and the ridge estimators for the FCNAR model through numerical experiments based on synthetic data, and the impact that the sample size $T$, the spline order $M$, the number of knots $K$, the size of the neighborhood in $W$ can have.

The data are generated from the following FCNAR$(1,1)$ model \begin{equation}\label{pe:1}
   \mathbb{X}_{t}=A(\mathbb{U}_{t})\mathbb{X}_{t-1}+B(\mathbb{U}_{t})W\mathbb{X}_{t-1}+\epsilon_t, 
\end{equation} 
where $a_i(u):=0.138+(0.316+0.982u)e^{-3.89u^2}$ and $b_i(u):=-0.437-(0.659+1.260u)e^{-3.89u^2}$, $i=1,2,\cdots,N$. The network size is fixed as $N=100$. Further, the $K$ knots  $(k_1,k_2,...k_{K})$ in the spline basis functions are evenly placed between the $1\%$ quantile and the $99\%$ quantile of the threshold variable $U_{it}$.

\subsection{Estimation of the Autoregressive and Network Effect Functions}

We examine the influence of the sample size $T$, the number of knots $K$, the spline order $M$ and the number of neighbors (bandwidth) of the network matrix $W$ on the estimation of the  $a(\cdot)$ and $b(\cdot)$ functions. For a grid $(k_1,k_2,...k_{200})$ of 200 points evenly placed  between $5\%$ quantile and $95\%$ quantile of the threshold variable $U_{it}$, $\hat{a}_i(u)$ and $\hat{b}_i(u)$ are calculated. The experiment is replicated 50 times, and  the $95\%$ quantile, the median and the $5\%$ quantile of the values of the functions at the 200 grid of knots are plotted and compared with the true functions.

The following simulation scenarios are considered:
\begin{enumerate}
    \item[A.1:] $\epsilon_{it}\sim N(0,1)$, $U_{it}\sim N(0,1)$ and $W$ is a banded matrix with bandwidth 2 neighbors. The order of the splines is set to 4, and $K=10$. Finally, the following 3 sample sizes are used: $T=200,400,3200$.
    \item[A.2:] The sample size is set to $T=400$ and 2400, and the spline order varies between $M=2,4,6$. Further, the error and threshold processes and $W$ are set as in scenario A.1.
    \item[A.3:] The order of the spline basis is set to $M=4$, while the number of knots varies according to $K=5,10,15$. The sample size, the error and threshold processes and the network matrix $W$ are as in scenario A.2.
    \item[A.4:] The order of the spline basis is set to $M=4$ with $K=10$ knots. The sample size and the error and threshold processes are set as in scenario A.2. The network matrix $W$ is banded and the bandwidth is set to $2,10,50$.
\end{enumerate}

\begin{figure} [H]
\centering
\subfigure[T=200]{
\includegraphics[scale=0.28]{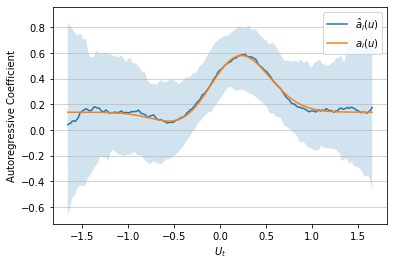}
}
\subfigure[T=800]{
\includegraphics[scale=0.28]{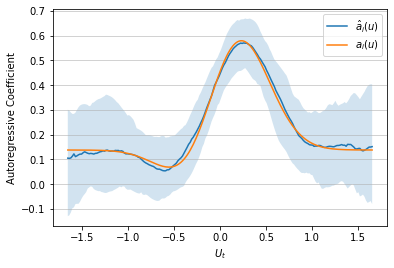}
} 
\subfigure[T=3200]{
\includegraphics[scale=0.28]{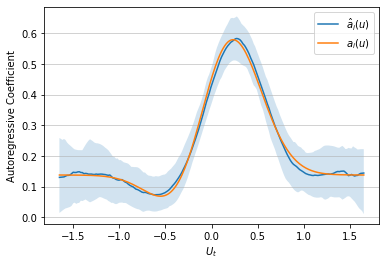}
}
\caption{$a_i(\cdot)$ and $\hat{a}_i(\cdot)$ autoregressive functions in scenario A.1.\label{a.1a}}
\subfigure[T=200]{
\includegraphics[scale=0.28]{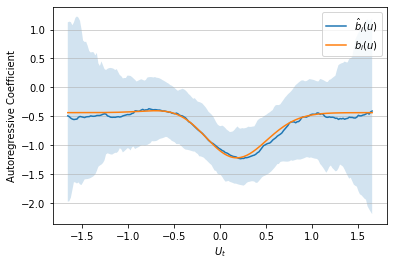}
} 
\subfigure[T=800]{
\includegraphics[scale=0.28]{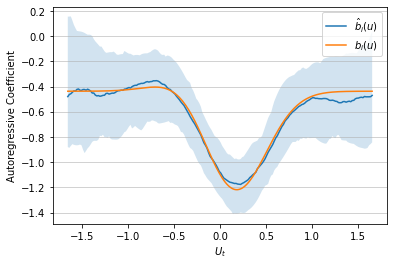}
} 
\subfigure[T=3200]{
\includegraphics[scale=0.28]{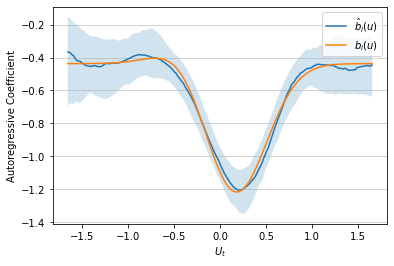}
} 
\caption{$b_i(\cdot)$ and $\hat{b}_i(\cdot)$ network effect functions in scenario A.1.\label{a.1b}}
\end{figure}

Figures \ref{a.1a} and \ref{a.1b} depict the true functions $a_i(\cdot)/b_i(\cdot)$ and their estimated counterparts $\hat{a}_{i}(\cdot)/\hat{b}_{i}(\cdot)$. It can be seen that the estimated functions get closer to the true coefficient functions as the sample size $T$ increases. In particular, for sample sizes larger than 800, the approximation is becoming fairly accurate.

Figures \ref{a.2a} and \ref{a.2b} in the supplementary material suggest that for exponential autoregressive and network effect functions, splines of low order ($M=2$) provide good estimates compared to higher order splines for $T=400$. This may be related to the limited sample size. As $T$ increases to 2400, the performance of splines with $M=4$ and 6 improve.

Figures \ref{a.3a} and \ref{a.3b} in the supplementary material show that the number of knots $K$ needs to be selected in accordance with the sample size $T$. Specifically, $K=10$ gives the best result, while for smaller $K$, there might be bias for the estimated functions, and for larger $K$ the estimated functions become unstable near the boundary. Finally, Figures \ref{a.4a} and \ref{a.4b} in the supplementary material confirm the robustness of the autoregressive function estimates over network matrices $W$ with different bandwidths (number of neighboring nodes included), while the network effect estimates become less accurate for more connected $W$ matrices. 

The results confirm that in empirical work wherein the sample size $T$ is given, the spline order $M$ and the number of knots $K$ need to be selected judiciously; for example, \citet{huang2004functional} propose using the AIC criterion for this task.



\subsection{Confidence Intervals and Coverage Probabilities}
Next, we examine the coverage of the confidence intervals based on the results of Theorem \ref{P:OLS}. The setting considered is as follows: the network size is set to $N=100$, the error process is distributed according to $\epsilon_{it}\sim N(0,1)$, $U_{it}\sim N(0,1)$ and $W$ is a banded matrix of bandwidth 2. A spline of order $M=4$ with $K=20$ knots is used to estimate the functions $a_i(\cdot)$ and $b_i(\cdot)$ while the sample size varies as 
$T=200,400,800,1600,3200$.
The experiment is replicated 100 times. 

\begin{figure} [H]
\centering
\subfigure[$U_{it}=-1.25$]{
 
\centering
\includegraphics[scale=0.32]{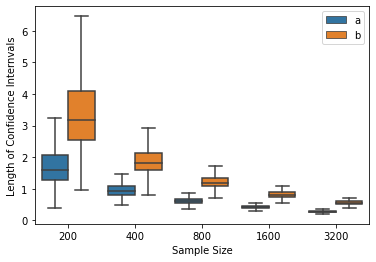}
 
}
\subfigure[$U_{it}=-1.25$]{
 
\centering
\includegraphics[scale=0.32]{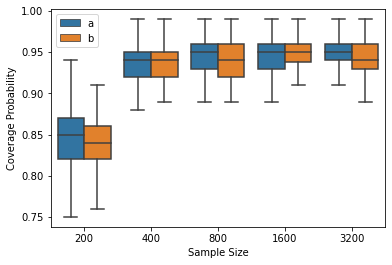}
 
} 
\subfigure[$U_{it}=-0.75$]{
 
\centering
\includegraphics[scale=0.32]{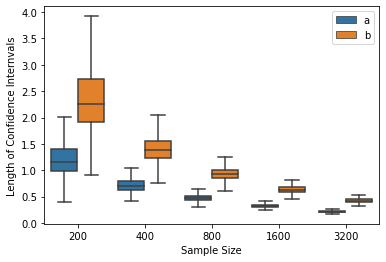}
 
}
\subfigure[$U_{it}=-0.75$]{
 
\centering
\includegraphics[scale=0.32]{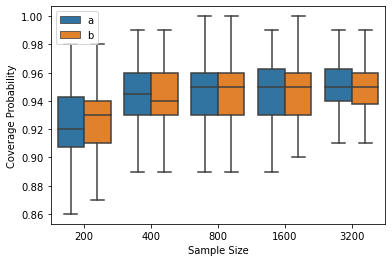}
 
}
\subfigure[$U_{it}=0$]{
 
\centering
\includegraphics[scale=0.32]{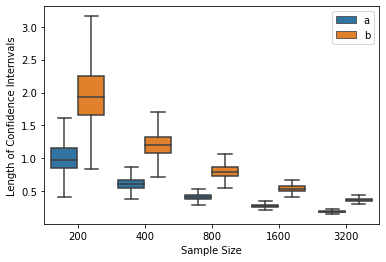}
 
}
\subfigure[$U_{it}=0$]{
 
\centering
\includegraphics[scale=0.32]{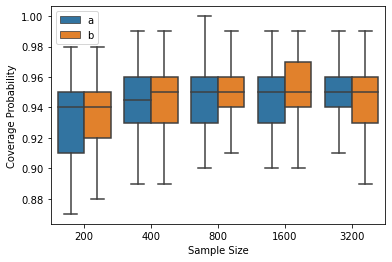}
 
}
\subfigure[$U_{it}=0.75$]{
 
\centering
\includegraphics[scale=0.32]{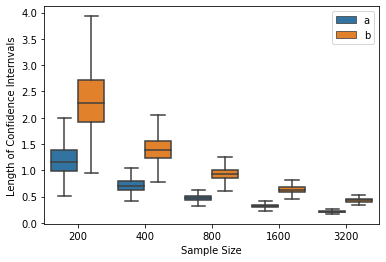}
 
}
\subfigure[$U_{it}=0.75$]{
 
\centering
\includegraphics[scale=0.32]{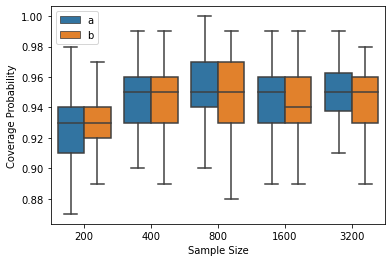}
 
}
\subfigure[$U_{it}=1.25$]{
 
\centering
\includegraphics[scale=0.32]{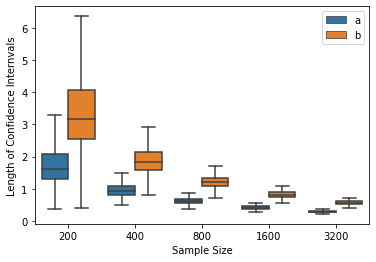}
 
}
\subfigure[$U_{it}=1.25$]{
 
\centering
\includegraphics[scale=0.32]{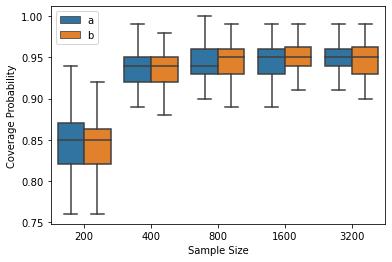}
 
}
\caption{CI and CP at different threshold of $U_{it}$\label{cicp}}
\end{figure}

Figure \ref{cicp} depicts the length of the confidence intervals (CI) and the coverage probabilities (CP) for different thresholds $U_{it}$. The $95\%$ CI is calculated using $CI_{a_i}=(\hat{a}_i(u)-z_{0.975}SE(\hat{a}_i(u)),\hat{a}_i(u)+z_{0.975}SE(\hat{a}_i(u)))$ and $CI_{b_i}=(\hat{b}_i(u)-z_{0.975}SE(\hat{b}_i(u)),\hat{b}_i(u)+z_{0.975}SE(\hat{b}_i(u)))$, where $SE(\hat{a}_i(u))$ and $SE(\hat{b}_i(u))$ are calculated according to Theorem \ref{P:OLS}.

It can be seen that the length of the CIs become smaller as the sample size $T$ increases. Further, for any fixed sample size, the length of the CIs is larger for points closer to the boundary of the domain of the threshold variable. On the other hand, with the exception of rather small sample sizes ($T=200$), the CP is close to the nominal level, especially for larger values of $T$. For points closer to the boundary, a larger sample size is required for the CP to be close to the nominal level.

 
 
 
 
 
 
 
 
 
 
 
 
 
 
 
 
 
 
 
 

\subsection{Prediction Performance}

\vspace{-0.5cm}

Next, we compare the predicted RMSE of different models, defined as:
\[\text{predicted RMSE}:=\sqrt{\frac{1}{NT}\sum\limits_{t=1}^T||\mathbb{X}_t-\mathbb{\hat{X}}_t||_F^2}.\]
The data is generated by \eqref{pe:1} with $N=100$. Two sets of functions $a_i(\cdot)$ and $b_i(\cdot)$ are tested:

\begin{enumerate}
    \item[B.1] $a_i(u):=0.138+(0.316+0.982u)e^{-3.89u^2}$ and $b_i(u):=-0.437-(0.659+1.260u)e^{-3.89u^2}$, $i=1,2,\cdots,N$.
    \item[B.2] $a_i(u):=0.3I(u\leq 1)-0.7I(u> 1)$ and $b_i(u):=-0.6I(u\leq 1)+0.2I(u> 1)$, $i=1,2,\cdots,N$.
    \item[B.3] $a_i(u):=0.138+(0.316+0.682u)e^{-0.5u^2}$ and $b_i(u):=-0.437-(0.259+0.560u)e^{-0.5u^2}$, $i=1,2,\cdots,N$.
\end{enumerate}

The first $T=1550$ time points of the data are used for estimating the model parameters, while the last $T=50$ time points to calibrate the prediction performance. We employ a FCNAR($1,1$), with an order-4 spline with $K=10$ knots to estimate the autoregressive $a_i(\cdot)$ and network effect $b_i(\cdot)$ parameters. Competing methods include a NAR(1,1) model (i.e, $a_i(\cdot)=a_i, b_i(\cdot)=b_i$, both constants) and a univariate autorgressive time series model with a single lag (as selected by the AIC criterion) applied to each node $i$ (AR(1)). It can be seen that FCNAR can capture the non-linearity in the data, and thus outperforms the competing models.

\begin{table}[H] 
\centering
\caption{predicted RMSE for different estimators\label{t:pe:prmse}}

\begin{tabular}{cccc}
  \hline
   &B.1&B.2&B.3\\ \hline
 FCNAR(1,1) & 1.015 & 1.022&1.017 \\ 
NAR(1,1) & 1.041 &1.104 &1.106 \\ 
AR(1) & 1.097 &1.135 &1.161  \\ \hline
\end{tabular}
\end{table}

\section{Applications of the FCNAR model to Real Data Sets} \label{sec:app}

\subsection{Application of the FCNAR model to Air Quality Indicators}
\label{sec:app-air-quality}

The proposed FCNAR model is employed to analyze 6 different air quality indicators, namely CO (carbon monoxide), O3 (ozone), SO2 (sulfur dioxide), NO2 (nitrogen dioxide), PM2.5 and PM10 (particulate matter in the air that are less than 2.5 and 10 micrometers in diameter, respectively). The data come from $N=346$ monitoring stations across China and are collected by the the China National Environmental Monitoring Center. After aggregating the original hourly data for the period 01/02/2015 to 09/05/2020 to a daily cadence and removing missing values, we end up with $T=1999$ daily measurements (observations). The first $1800$ days are used to estimate the FCNAR model parameters, while the last $200$ days are used for testing purposes. The $AIC=\log |\hat{\Sigma}_\epsilon|+\frac{2(K+M)(q_1+q_2)}{T}$ information criterion is used to select the number of knots, the lag order, the lag of the threshold variable (which corresponds to the response variable) and the spline order. The selected spline order is $M=3$, the number of knots $K=5$ and $\mathbb{U}_{t}:=\mathbb{X}_{t-4}$. 

\begin{figure} [H]
\centering
\subfigure[CO]{

\includegraphics[scale=0.3]{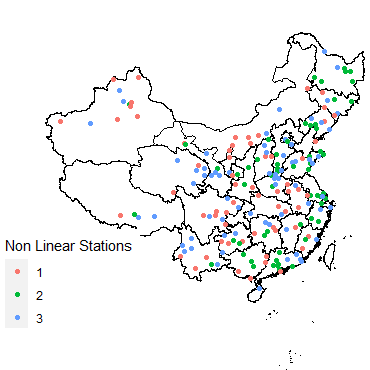}

}
\subfigure[O3]{

\includegraphics[scale=0.3]{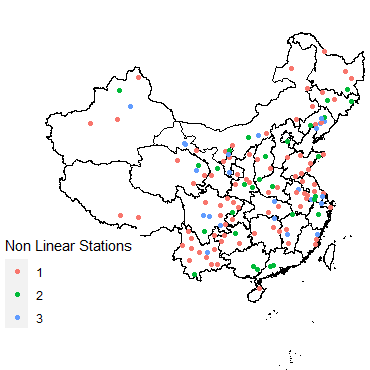}

} 
\subfigure[SO2]{

\includegraphics[scale=0.3]{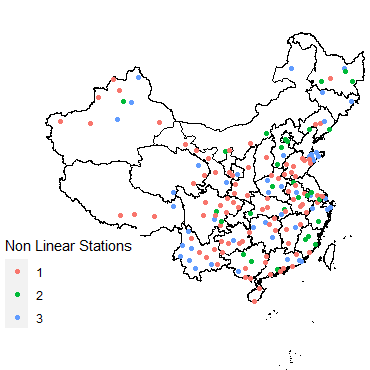}

} 
\subfigure[NO2]{

\includegraphics[scale=0.3]{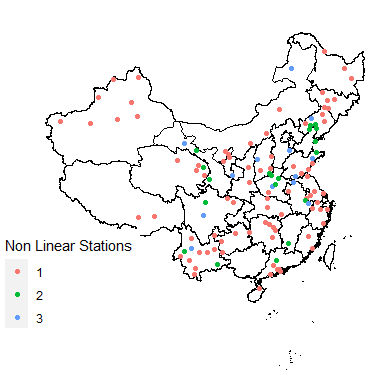}

} 
\subfigure[PM 2.5]{

\includegraphics[scale=0.3]{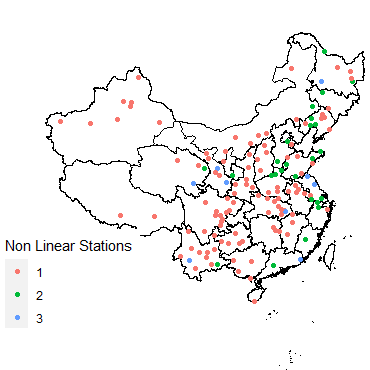}

} 
\subfigure[PM 10]{

\includegraphics[scale=0.3]{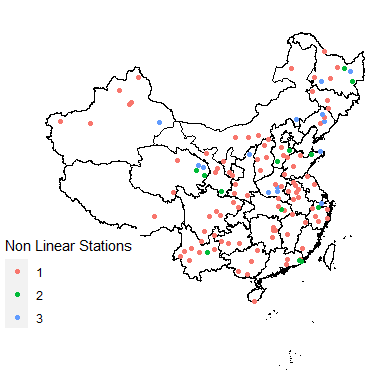}

} 
\caption{Stations with Non-linear $a(\cdot)$ (red), $b(\cdot)$ (green) and both $a(\cdot)$ and $b(\cdot)$ (blue)\label{non-linear}}
\end{figure}

The network matrix $W$ is obtained as follows: let $D_{ij}$ be the spatial distance between two monitoring stations, then the $ij$-th element of $W$ is defined as:

The F-test (for details see Section \ref{sec:testing}) for the following two null hypotheses
\[H_0:a_{i2}=a_{i3}=\cdots=a_{i8}=0,\ or\]
\[H_0:b_{i2}=b_{i3}=\cdots=b_{i8}=0\]
is used to test, if each node exhibits a non-linear autoregressive or network effect. The location of the monitoring stations that exhibit such non-linear stations are depicted in Figure \ref{non-linear}; specifically, red points represent nodes with non-linear autoregressive coefficients, green points represent nodes with non-linear network coefficients, and blue points represent nodes with both non-linear autoregressive and network coefficients. Figure \ref{non-linear} indicates that for most of the stations, there exist either non-linear autoregressive or network coefficients. Hence, the proposed FCNAR model that accounts for heterogeneity proves useful for modeling these 6 air quality indicators.

Next, we assess the predictive performance of the FCNAR vis-a-vis selected competitors. Specifically, the predicted RMSE of the ridge based FCNAR(1,1) model is compared to that of a NAR(1,1) model (with linear autoregressive and network effects), a univariate AR(1) model applied to each node, as well as a NAR(1,1) model with homogeneous coefficients (i.e, $a_i=a, b_i=b$ for all $i$ nodes). The tuning parameter $\lambda$ for the ridge based FCNAR(1,1) model is selected by cross-validation. The predicted RMSE of the competing models is given in Table \ref{t:prmse}. Overall, FCNAR(1,1) performs the best across all air quality indicators, with the linear NAR(1,1) following behind. 
\begin{table}[H] 
\centering
\caption{predicted RMSE for different estimators\label{t:prmse}}
\resizebox{\columnwidth}{!}{
\begin{tabular}{c|c|c|c|c|c|c}
 
  \hline
  &PM10&PM2.5&SO2&NO2&O3&CO\\ \hline
 FCNAR(1,1) &  0.93425& 0.87453&0.71338&0.95353& 0.83950&0.66762 \\ \hline
NAR(1,1) & 0.93575&0.87645&0.71433&0.95380&0.84014&0.66863  \\ \hline
AR(1) & 0.94832&0.88824&0.72411&0.96565&0.84507&0.67376   \\ \hline
NAR with $A=aI$ and $B=bI$ & 0.94060&0.88210&0.72097&0.96192&0.84420&0.67306   \\ \hline
\end{tabular}
}
\end{table}

We also compare the relative predicted RMSE improvement of the FCNAR(1,1) model compared to its FAR component (i.e, assuming all $b_i(\cdot)=0$), and the results are depicted in the next Figure. It can be seen that for the coastal area, the functional network effects are more significant, bringing $4\%$-$6\%$ improvement in predictive performance.
\begin{figure} [H]
\centering
\subfigure[CO]{

\includegraphics[scale=0.33]{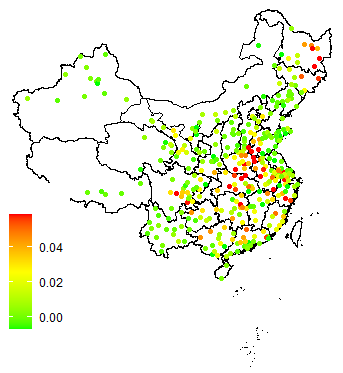}

}
\subfigure[O3]{

\includegraphics[scale=0.33]{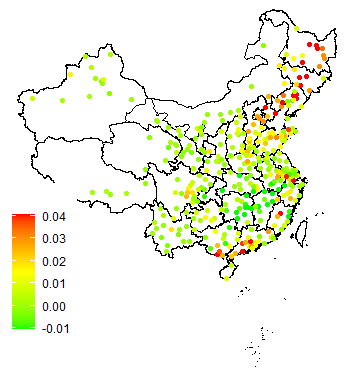}

} 
\subfigure[SO2]{

\includegraphics[scale=0.33]{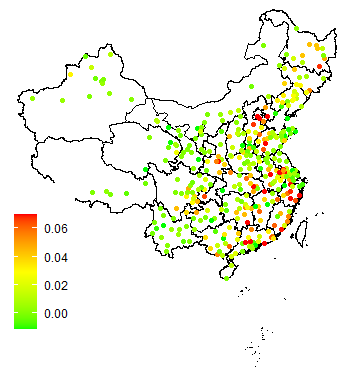}

} 
\subfigure[NO2]{

\includegraphics[scale=0.33]{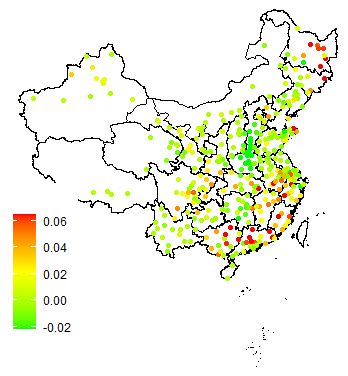}

} 
\subfigure[PM 2.5]{

\includegraphics[scale=0.33]{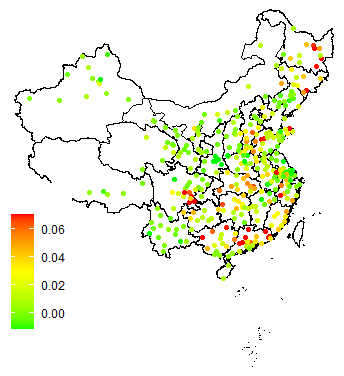}

} 
\subfigure[PM 10]{

\includegraphics[scale=0.33]{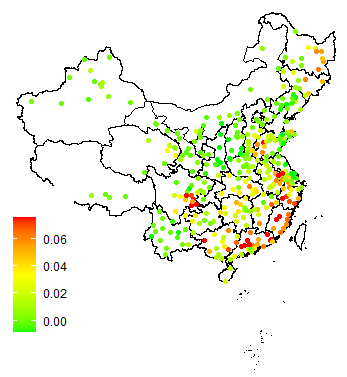}

} 
\caption{Relative predicted RMSE improvement compared with pure FAR component}
\end{figure}

Based on the previous finding, we focus on the predictive performance for the stations in Southeast China - the provinces of Fujian and Guangdong. 
Table \ref{t:prmse:se} compares the predicted RMSE of the FCNAR(1,1) model with the FAR(1) and AR(1) model for the respective stations. Overall, compared to a pure FAR model, there's a $3\%$-$4\%$ improvement in predicted RMSE for the stations in the Southeast by incorporating information from neighboring stations through the FCNAR model.
\begin{table}[H] 
\centering
\caption{predicted RMSE for different estimators (southeast)\label{t:prmse:se}}
\resizebox{\columnwidth}{!}{
\begin{tabular}{c|c|c|c|c|c|c}
 
  \hline
  &PM10&PM2.5&SO2&NO2&O3&CO\\ \hline
 FCNAR(1,1) &  0.84700 & 0.87803& 0.64093&0.87323& 0.82152&0.69485 \\ \hline
AR(1) & 0.88033 &0.90849&0.65938&0.89514&0.82630&0.70224   \\ \hline
FAR(1) & 0.87931 &0.90807&0.65787&0.89576&0.82657&0.70220   \\ \hline
\end{tabular}
}
\end{table}

For a grid $(k_1,k_2,...k_{200})$ of 200 points evenly placed between $1\%$ quantile and $99\%$ quantile of the threshold variable $U_{it}$, $\hat{a}_i(u)$ and $\hat{b}_i(u)$ for $i=1,2,\cdots,346$ are calculated. Based on the 200 $\hat{a}_i(u)$ values and 200 $\hat{b}_i(u)$ values, a total of 400 values for each station, k-means is utilized to partition 346 stations into three clusters. It can be seen that there's a significant geometric pattern of the functional coefficients, from the southeast to the northwest.

\begin{figure} [H]
\centering
\subfigure[CO]{

\includegraphics[scale=0.3]{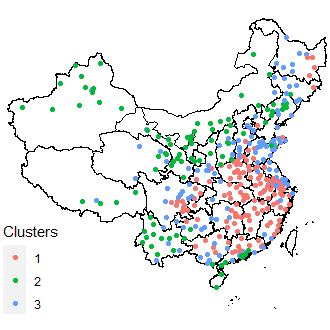}

}
\subfigure[O3]{

\includegraphics[scale=0.3]{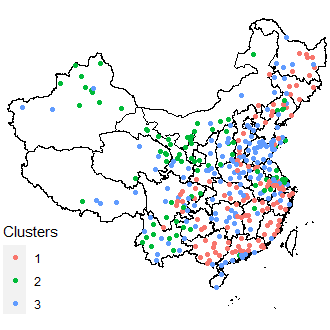}

} 
\subfigure[SO2]{

\includegraphics[scale=0.3]{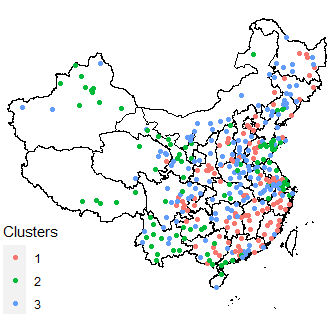}

} 
\subfigure[NO2]{

\includegraphics[scale=0.3]{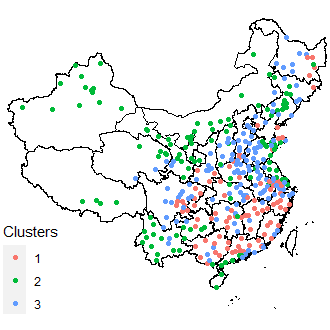}

} 
\subfigure[PM 2.5]{

\includegraphics[scale=0.3]{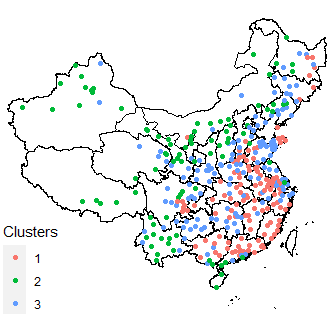}

} 
\subfigure[PM 10]{

\includegraphics[scale=0.3]{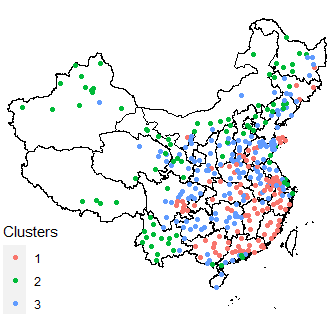}

} 
\caption{Clusters}
\end{figure}















\subsection{Application to Florida County Level COVID-19 Data}\label{sec:app-covid}

The COVID-19 data used in this study are obtained from \citet{covid} website. The analysis focuses on the 67 counties in the state of Florida and the raw data include both cases and deaths, as reported by state and local health departments and compiled by the newspaper. The network matrix $W$ is calculated from the county adjacency file obtained from the \cite{adj}. The network matrix is normalized with each row summing up to 1.

The data we focused on is the daily count of \textit{new} COVID-19 cases for the $N=67$ Florida counties, covering the period from March 2, 2020 (when the Center for Disease Control reported the first Covid-19 cases in the state of Florida), to Dec 31, 2022, for a total $T=1036$ observations. The first 900 days are used as the training set, up to 08/17/2022, and the last 136 days are used as the test set. 

The $AIC=\log |\hat{\Sigma}_\epsilon|+\frac{2(K+M)(q_1+q_2)}{T}$ information criterion is used to select the number of knots, the lag order, the lag of the threshold variable (which corresponds to the response variable) and the spline order. The model selected is FCNAR$(2,2)$ with spline order $M=3$ and $K=3$ knots, with $\mathbb{U}_{t}:=\mathbb{X}_{t-1}$. 

\begin{figure} [H]
\centering
\subfigure[$a_i(\cdot)$]{
\includegraphics[scale=0.3]{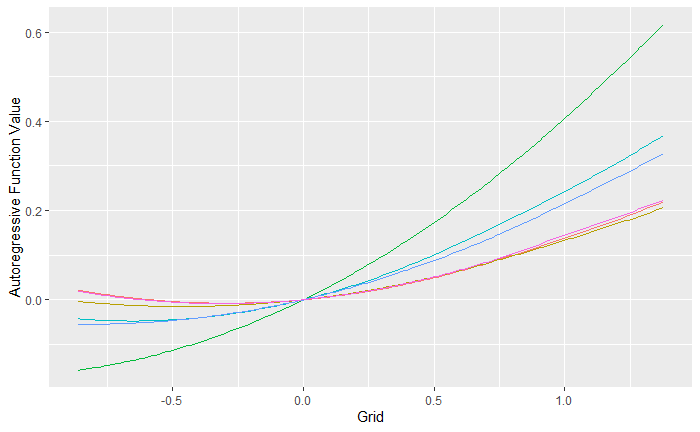}
}
\subfigure[$b_i(\cdot)$]{
\includegraphics[scale=0.3]{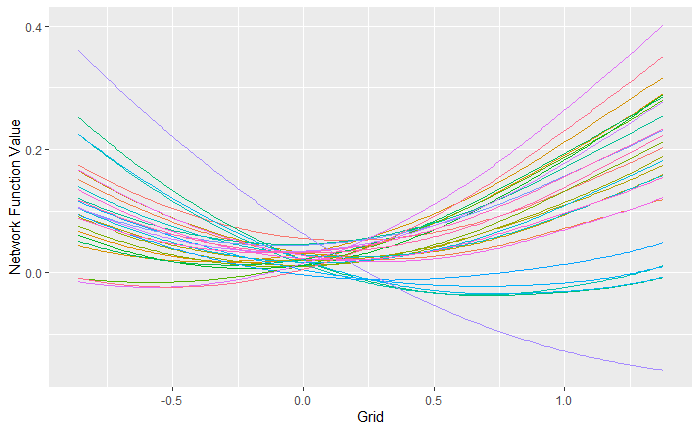}
} 
\caption{Plot of autoregressive ($a_i(\cdot)$) and network ($b_i(\cdot)$) effects with strong nonlinear patterns\label{fig:functions-of-covid}}
\end{figure}

\begin{figure} [H]
\centering
\subfigure[$a(\cdot)$]{
\includegraphics[scale=0.3]{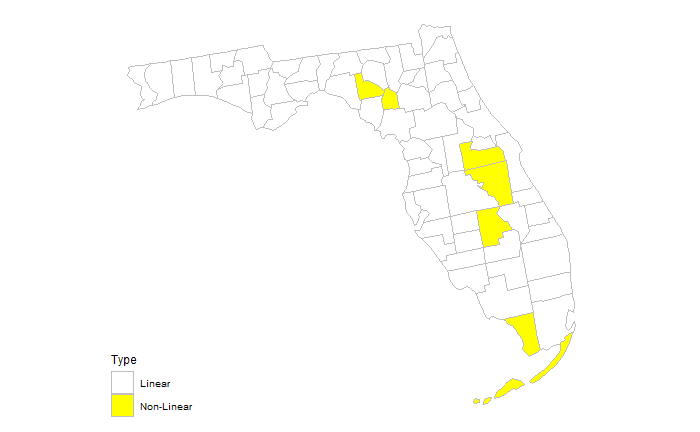}
}
\subfigure[$b(\cdot)$]{
\includegraphics[scale=0.3]{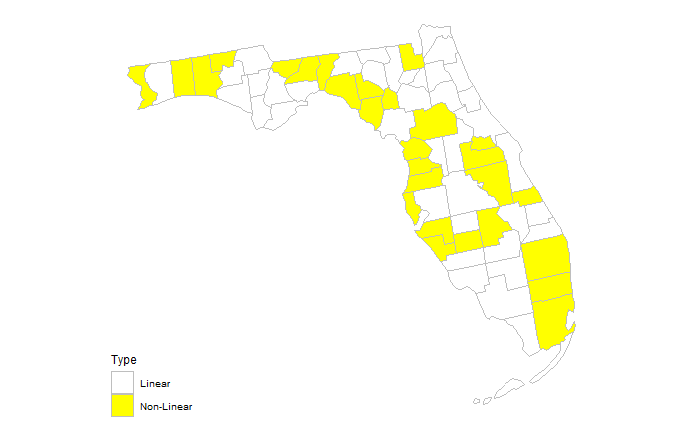}
} 
\caption{Counties with Non-linear autoregressive ($a(\cdot)_i$) and network ($b_i(\cdot)$) effects. \label{fig:non-linear-covid}}
\end{figure}

Plots of selected autoregressive $a_i(\cdot)$ and network effect $b_i(\cdot)$ functions that exhibit strong nonlinear patterns are depicted in Figure \ref{fig:functions-of-covid}. It can be seen that the larger the lag-value, the larger the autoregressive/network effect. Further, as gleaned from Figure \ref{fig:non-linear-covid} mostly coastal counties exhibit nonlinear network effects

\begin{figure} [H]
\centering
\subfigure[\label{f1}]{
\includegraphics[scale=0.21]{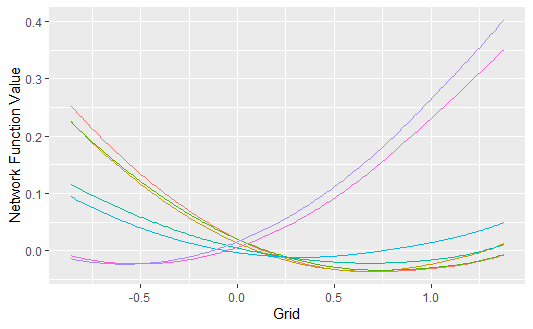}
}
\subfigure[\label{f2}]{
\includegraphics[scale=0.21]{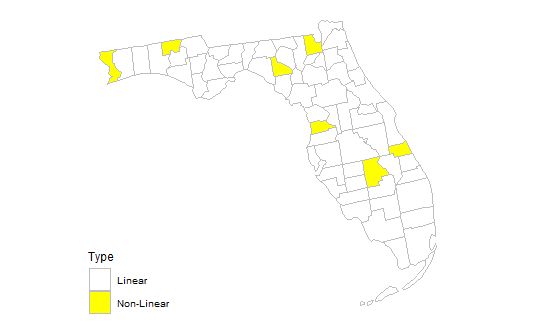}
}
\subfigure[\label{f3}]{
\includegraphics[scale=0.21]{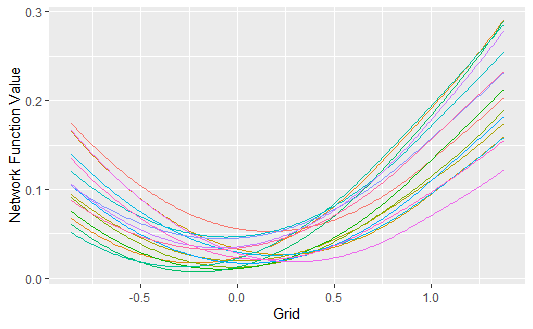}
}
\subfigure[\label{f4}]{
\includegraphics[scale=0.21]{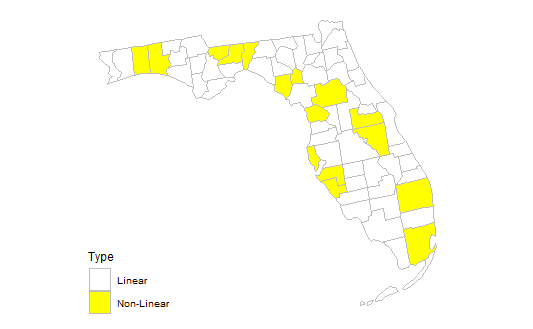}
} 
\caption{The left panels (\ref{f1}, \ref{f2}) and the right panels (\ref{f3}, \ref{f4}) depict $b_i(\cdot)$ of the two patterns with their corresponding locations\label{fig:g1-of-covid}}

\end{figure}

A closer examination of the network effect functions ($b_i(\cdot)$) shows two distinct patterns, as shown in Figure \ref{fig:g1-of-covid}. The first is increasing/decreasing (Figure \ref{f1} and the counties associated with it are shown in Figure \ref{f2}, while the other is a quadratic one (Figure \ref{f3}) and the corresponding counties depicted in Figure \ref{f4}.

Next, the predictive performance of the developed FCNAR model is compared with a standard NAR model and a node-wise AR model.
\begin{table}[H] 
\centering
\caption{predicted RMSE for different estimators\label{t:prmse2}}
\resizebox{\columnwidth}{!}{
\begin{tabular}{c|c|c|c|c}
 
  \hline
 Model &FCNAR(2,2)&FCNAR(2,2) with ridge&NAR(2,2)&AR(2)\\ \hline
predicted RMSE  &  0.6791& 0.6777&0.7521&0.7781 \\ \hline

\end{tabular}
}
\end{table}
It can be seen that the proposed FCNAR model clearly outperforms the standard NAR one by over 10\% in terms of predicted RMSE. The latter also exhibits a small advantage over a univariate AR model fitted to each county separately. For this data set, the added flexibility encompassed in FCNAR contributes to the improved performance.

\section{Concluding Remarks}\label{sec:con}

The paper introduces the FCNAR modeling framework that extends both the linear (heterogeneous) NAR model of \citet{yin2021general} by accommodating nonlinear autoregressive and network effects and the univariate FAR model of \citet{chen1993functional}
to a multivariate setting characterized by an underlying network structure. 
A sufficient condition for the FCNAR process to be stable is provided and we also present least squares and ridge estimators for the functional model parameters that are expressed in spline bases. We establish their asymptotic properties that enable testing of a number of interesting hypotheses, including (i) whether the autoregressive and/or network effects are linear or nonlinear, and (ii) whether the functional network effects components are zero or not, thus reducing the FCNAR to a collection of unrelated FAR models for each node. Experiments on synthetic data assess the performance of the FCNAR model for different sample sizes, and selection of spline basis parameters (order, number of knots). Finally, an application to 6 air quality indicators collected across a network of monitoring stations shows that the data exhibit both nonlinear autoregressive and network effects, as well as heterogeneity across the stations.



\vspace{-0.6cm}

\section*{Supplementary Material}

\vspace{-0.5cm}

The Supplementary material contains statements of auxiliary technical results, the proofs of all Lemmas and Theorems, additional simulation results showcasing the sensitivity of the model parameter estimates to the specification of the network weight matrix $W$ and analysis of an additional data set on wind speeds.

\par



\vspace{-0.4cm}

\bibhang=1.7pc
\bibsep=2pt
\fontsize{9}{14pt plus.8pt minus .6pt}\selectfont
\renewcommand\bibname{\large \bf References}
\expandafter\ifx\csname
natexlab\endcsname\relax\def\natexlab#1{#1}\fi
\expandafter\ifx\csname url\endcsname\relax
  \def\url#1{\texttt{#1}}\fi
\expandafter\ifx\csname urlprefix\endcsname\relax\def\urlprefix{URL}\fi

  \bibliographystyle{chicago}      
  \bibliography{ref}   

\vskip .65cm
\noindent
University of Florida, 
E-mail: hyin@ufl.edu
\vskip 2pt

\noindent
George Mason University,
E-mail: asafikha@gmu.edu
\vskip 2pt

\noindent
University of Florida,
E-mail: gmichail@ucla.edu

\end{document}